\documentclass[aps,prd,onecolumn, groupedaddress, showpacs, 11pt, tightenlines]{revtex4} 

\usepackage{graphicx}
\usepackage{amsmath}  

\begin{document}

\title{Trapping effects in inflation: blue spectrum at small scales}

\author{Edgar Bugaev}
\email[e-mail: ]{bugaev@pcbai10.inr.ruhep.ru}

\author{Peter Klimai}
\email[e-mail: ]{pklimai@gmail.com}
\affiliation{Institute for Nuclear Research, Russian Academy of
Sciences, 60th October Anniversary Prospect 7a, 117312 Moscow, Russia}




\begin{abstract}
We consider the inflationary model in which the inflaton $\phi$
couples to another scalar field $\chi$ via the interaction
$g^2(\phi-\phi_0)^2\chi^2$ with a small coupling constant $g$
($g^2 \sim 10^{-7}$). We assume that there is a sequence of
``trapping points'' $\phi_{0i}$ along the inflationary trajectory
where particles of $\chi$-field become massless and are rather
effectively produced. We calculate the power spectrum of inflaton
field fluctuations originated from a backreaction of $\chi$-particles
produced, using the Schwinger's ``in-in'' formalism. We show that
the primary curvature power spectrum produced by these backreaction
effects is blue, which leads to a strong overproduction of
primordial black holes (PBHs) in subsequent radiation era.
\end{abstract}



\maketitle

\section{Introduction}
\label{sec-intro}

In many works of recent time a large class of inflationary
models has been considered in which a generation of
inflaton field fluctuations in a course of inflation is
implemented (partly, at least) through an interaction between
inflaton and other quantum fields. In particular,
in \cite{Kofman:1986wm, Chung:1999ve, Kofman:2004yc,
Romano:2008rr, Green:2009ds, Barnaby:2009mc, Barnaby:2010ke,
Lee:2011fj, Wu:2006xp}
the interaction of the type
\begin{equation}
\label{calLint}
{\cal L}_{\rm int} = -\frac{g^2}{2} ( \phi - \phi_0  )^2 \chi^2
\end{equation}
has been thoroughly exploited. Here, $\phi$ is the inflaton field,
and $\chi$ is some other scalar field. It has been shown that
the interaction of this type leads to a particle creation during inflation,
because when $\phi$, in a process of slow-roll, comes nearer to
$\phi_0$, the particles of $\chi$ field become massless and are effectively
produced. During the short time interval of $\chi$-particle
production, a feature in the power spectrum of scalar curvature
fluctuations is generated. In early works \cite{Chung:1999ve}
(and, also, in \cite{Romano:2008rr}) the process has been studied
in the mean-field approximation (i.e., the variance
$\langle \chi^2\rangle$, solely, has been used to quantify the
back reaction of $\chi$-particles, produced during inflation, on
the inflaton field). In subsequent works the calculations
were refined, going beyond the mean-field treatment. Methods
used for the calculation include
{\it i)} the analytic description of particle creation with
the coupling (\ref{calLint}) developed in the theory of
preheating after inflation \cite{Kofman:1994rk, Kofman:1997yn},
{\it ii)} cosmological perturbation theory for the field
equations (see, e.g., \cite{Seery:2008qj}) (the method is
used in \cite{Barnaby:2009mc, Barnaby:2010ke}) and
{\it iii)} Schwinger's ``in-in''-formalism generalized
to compute cosmological perturbations (this method is
used in \cite{Lee:2011fj, Wu:2006xp}).

In the present work we calculate the power spectrum of
inflaton field fluctuations originated from a back reaction
of $\chi$-particles produced during inflation, via
the coupling (\ref{calLint}), on the inflaton field.
We studied, in contrast with \cite{Green:2009ds,
Barnaby:2009mc, Barnaby:2010ke}, only the case of weak
coupling, $g^2 \sim 10^{-7}$, and, exclusively, the
region of small scales, $k/H \gg 1$ ($k$ is the comoving
wave number, $H$ is the Hubble parameter during inflation,
scale factor $a$ is equal to $1$ at the initial time of
the inflation era). If $g^2$ is so small, the ``trapped
inflation'' scenario \cite{Kofman:2004yc, Green:2009ds}
is ineffective \cite{Barnaby:2009mc} but just for this region
of $g^2$-values the interesting phenomenon was predicted
in \cite{Lee:2011fj}. Namely, authors of \cite{Lee:2011fj}
(see also \cite{Liu:2014ifa})
argued that in a case of closely-spaced trapping points,
i.e., if we have a sequence of points $\phi_{0,\; i}$
($i = 1, ... , n$) where particles $\chi_i$ become massless,
the total (accumulative) power spectrum of inflaton fluctuations
at small scales would be blue.  This conclusion is important
because such small scale fluctuations might effect
primordial black hole (PBH) formation when the fluctuations
will cross horizon inside after an end of inflation.

The approach used in the present work is, technically, rather
similar with those of refs \cite{Morikawa:1989xz, HZ, Matarrese:2003ye}.
In these works the original scenario
of stochastic inflation \cite{AAS} had been reformulated using functional
methods. In \cite{Morikawa:1989xz, HZ, Matarrese:2003ye},
as in \cite{AAS}, the Fourier components of the
inflaton field, as a whole, are splitted into long-wavelength modes
(with wavelengths $a/k$ longer than the horizon scale $H^{-1}$) and
short-wavelength ones. This splitting is performed now at the action
level, and short-wavelengths are integrated out via a path integral
over the sub-horizon part of the whole field. The long-wavelength modes
of the field are assumed to be classical, by itself, but long-short
couplings (due to a time dependence of the window function, in particular)
yield the semi-classical correction to their equation of motion. In
our case, the  key difference with this approach is that, instead
of the short-wavelength part of the inflaton field, we have the
independent field $\chi$ interacting with the inflaton field. Further,
in our case the inflaton field is not coarse-grained by the splitting
of its own modes; the semi-classical corrections to the equation of
motion for modes of the inflaton field arise due to loops of $\chi$-field
only. This approach, in application to calculations of the power spectrum
of inflaton fluctuations, is justified if the correction to this spectrum
from these loops is formed at around a time when the corresponding mode
exits horizon.

For concrete calculations we use in this paper the closed-time-path
(CTP) functional formalism of Schwinger and Keldysh \cite{SK}.
The application of this formalism to cosmological problems
had been suggested in pioneering works by Calzetta and Hu
\cite{CH} and Jordan \cite{Jordan:1986ug}. It is well known
(see, e.g., \cite{Calzetta:1993qe, Hu:1994iw}) that this
formalism is especially useful for studying cosmological
backreaction problems. Most importantly, this approach
which operates with ``in-in'' effective action yields
the real and causal equations of motion describing a time evolution
of the system (inflaton scalar field, in our case). Furthermore,
the CTP functional formalism is closely related to the influence
functional formalism of Feynman and Vernon \cite{FV} because
in both methods the full quantum system can be divided in two
parts: the distinguished subsystem (the ``open system'', or, simply,
the ``system'') and the remaining degrees of freedom (the
``environment'', in our case, this is the $\chi$ field).
We are interested in the state of the system as influenced by the
overall effect of the environment, so, the environmental degrees
of freedom must be integrated out. It is easy to verify that
integrating out these variables in a CTP path integral, the
generating CTP functional can be expressed in terms of the
influence functional of Feynman and Vernon. Correspondingly, in
the semiclassical approximation, the effective CTP action
is expressed through the influence action.

The influence of the environment on the (open) system is, by
definition, the backreaction effect. The influence action is,
in general, complex; its real part contains the dissipational
kernel which yields the dissipative terms in the effective
equations of motion. The imaginary part contains the noise kernel
accounting the fluctuations induced on the system through
its coupling to the environment (we use here the terminology
of \cite{Calzetta:1993qe, Hu:1994iw}).

As is well known, the dynamical evolution of the (open) system
is not deterministic, even in the semiclassical approximation,
it is, in general, stochastic \cite{Calzetta:1993qe, Hu:1994iw}.
This is well illustrated by the quantum Brownian model
\cite{Caldeira:1982iu, Grabert:1988yt}, in which the
Feynman-Vernon idea of a stochastic force from the environment
acting on the system, had been exploited. In this approach, the
time evolution of the system degrees of freedom is described
by the Langevin equation.

The kernels of the influence action of two interacting quantum
fields in de Sitter space had been found in pioneering work
by Hu, Paz and Zhang \cite{Hu:1992ig}. Later, the similar
influence actions and functionals had been considered in
\cite{Gleiser:1993ea, Berera:1998gx} (in Minkowski space),
in \cite{Yokoyama:2004pf}, in warm inflation models (see
\cite{Ramos:2013nsa} and references therein), and in works on
effective field theory \cite{Boyanovsky}.

Naturally, the backreaction of the environment on the open
system cannot be too strong, so as to make meaningless
the separation scheme. In our case the open system is the
inflaton field, i.e., we consider the scenario of effective
single field inflation. The backreaction is significant during
the time $\Delta t$ when the $\chi$-field is light, so the
condition for a reasonable separation is $H\Delta t \lesssim 1$ \cite{Barnaby:2010ke},
where $H$ is the Hubble parameter during inflation.

If the environmental scalar field is minimally coupled to
gravity and nearly massless (and this is just the case considered
in the present work), the additional problem connected with
infrared divergences, arises. The kernels of the influence
action are expressed through the momentum integrals. In particular,
an one-loop contribution from integrating out the environment
field is given by momentum integrals over product of four mode
functions of this field. These integrals are divergent in the
massless limit (see, e.g., \cite{Boyanovsky:2012}).

The main feature of massless, minimally coupled (MMC) scalar fields
is the absence of normalizable de Sitter invariant states,
in other words, there is no de Sitter-invariant Fock vacuum state
\cite{Allen:1985ux, Allen:1987tz}. In particular, the Bunch-Davies
vacuum breaks the de Sitter invariance, when $m = \xi = 0$.
The consequences of this breaking important for us are:
{\it i)} the mean squared fluctuations of MMC field grow linearly
with time during inflation \cite{BDetal} and
{\it ii)} the de Sitter invariant two-point function (propagator)
becomes infrared divergent in the limit $m=\xi=0$ \cite{Ford:1977in}.

For a regularization of the infrared divergence we use
in the mode expansion of the $\chi$ field the comoving infrared cutoff
$\Lambda$ \cite{Lyth:2007jh, Bartolo:2007ti}. We put $\Lambda = H$;
this cutoff value is very natural \cite{Prokopec:2007ak, Xue:2011hm, Larjo:2011uh}
if we want to set initial conditions for all modes at the beginning
of inflation because the physics inside of the initial Hubble radius
$H^{-1}$ cannot determine the initial conditions for super-Hubble modes
(clearly, there is no causal process for preparing the initial state
in a space box having a size larger than horizon).

As a result of our work, we obtain a qualitative confirmation
of the main conclusion of \cite{Lee:2011fj} about the blueness
of the power spectrum at small scales. Our calculation,
however, differs from those of \cite{Lee:2011fj} by some
important details. In particular, we were not able to derive
and use their basic formula for the power spectrum.
Further, our results are quite sensitive to a value of the
infrared cut-off parameter $\Lambda$ (see below), while
we do not see something similar in formulas of \cite{Lee:2011fj};
in particular, their final expression for the noise-driven power spectrum does 
not contain the infrared cut-off explicitly.

The plan of the paper is as follows.
In the second section we formulate our theoretical approach and
obtain the equation for the amplitude of inflaton field fluctuations
needed for a power spectrum calculation.
The third section contains derivation of the power spectrum formula.
In the last section we present the results of our power spectrum calculations
and main conclusions about constraints on the parameters of the model
following from PBH overproduction predictions.

\section{Effective action and Langevin equation}

As is stated in the Introduction, we use the 
closed-time-path (CTP) formalism of Schwinger and Keldysh \cite{SK}
and the influence functional method of Feynman and Vernon \cite{FV}.
Our Lagrangian is
\begin{eqnarray}
\label{calLders}
{\cal L} = \frac{1}{2}g^{\mu\nu} \partial_\mu \phi \partial_\nu \phi +
\frac{1}{2} g^{\mu\nu} \partial_\mu \chi \partial_\nu \chi - V(\phi) -
\frac{g^2}{2}\phi^2\chi^2 \; , \\
\phi = \Phi - \Phi_0  \; . \qquad \qquad \qquad  \qquad \nonumber
\end{eqnarray}
The splitting between the system and the environment in our case is as
follows: the system sector contains all the modes of inflaton field
$\phi$, and the environment contains the modes of massless scalar
$\chi$-field with physical wavelengths shorter than the critical
length $\lambda_c = 2\pi/\Lambda$ at the initial conformal time $\eta_i$.
We set $a(\eta_i)=1$, so a physical length
$\lambda_{\rm phys} = a(\eta) \lambda$ coincides with the comoving
length $\lambda$ at $\eta = \eta_i$. We approximate the space-time during inflation
by a de Sitter metric,
\begin{equation}
ds^2 = a^2(\eta) (d\eta^2 - d \vec x ^2 ), \qquad
a(\eta) = \frac{-1}{H\eta} \;.
\end{equation}

The generating CTP functional is defined by introducing sources for
the $\phi$ field modes only:
\begin{eqnarray}
\label{eiWctp}
e^{i W_{CTP} [J^+, J^-] } = \int d\phi_f d\phi_i^+ d\phi_i^- \times
\qquad \qquad \qquad \qquad \qquad \qquad \qquad \qquad \qquad \qquad \\ \times \nonumber
\int\limits_{\phi_i^+}^{\phi_f} {\cal D} \phi^+
\int\limits_{\phi_i^-}^{\phi_f} {\cal D} \phi^-
e^{i(S_0[\phi^+] - S_0[\phi^-] + J^+\phi^+ - J^-\phi^- + S_{IF}[\phi^+, \phi^-; \infty])} \cdot
\rho_\phi(\phi_i^+, \phi_i^-; t_i) \; .
\end{eqnarray}
Here, $\rho_\phi$ is the initial density matrix for the $\phi$ field.

The expressions for free field actions and for the action describing the interaction
of the fields are given by
\begin{eqnarray}
\label{Sint}
S_{int}[\phi, \chi] = -\frac{g^2}{2}\int d^4 x a^4(\eta) \phi^2 \chi^2 \; ,
\end{eqnarray}
\begin{eqnarray}
\label{S0phi}
S_0[\phi] = \int d^4 x a^2(\eta) \left[
  \frac{\phi'^2}{2} - \frac{(\nabla \phi)^2}{2} - a^2(\eta) V(\phi)
\right] \;,
\end{eqnarray}
\begin{eqnarray}
\label{S0chi}
S_0[\chi] = \int d^4 x a^2(\eta) \left[
  \frac{\chi'^2}{2} - \frac{(\nabla \chi)^2}{2}
\right] \; .
\end{eqnarray}

The influence action is expressed by the formula:
\begin{eqnarray}
\label{eiSIF}
e^{i S_{IF}[\phi^+, \phi^-; t_f] } = \int d\chi_f d\chi_i^+ d\chi_i^- \times
\qquad \qquad \qquad \qquad \qquad \qquad \qquad \qquad
\qquad \qquad \qquad \qquad \qquad \\ \times \nonumber
\int\limits_{\chi_i^+}^{\chi_f} {\cal D} \chi^+
\int\limits_{\chi_i^-}^{\chi_f} {\cal D} \chi^-
e^{i(S_0[\chi^+] + S_{int}[\phi^+, \chi^+] - S_0[\chi^-] - S_{int}[\phi^-, \chi^-])} \cdot
\rho_\chi(\chi_i^+, \chi_i^-; t_i),
\end{eqnarray}
where $\rho_\chi$ is the initial density matrix for the $\chi$ field.
The CTP or in-in effective action containing all quantal
corrections to the field expectation value is given by
\begin{eqnarray}
\label{Gampm}
\Gamma(\bar\phi^+, \bar\phi^-) = W_{CTP}[J^+, J^-] - J^+\bar\phi^+ +
J^-\bar\phi^- \;,
\end{eqnarray}
\begin{eqnarray}
\label{phipmdef}
\bar\phi^{\pm} = \frac{\delta W_{CTP}[J^+, J^-]}{\delta J^{\pm}} \; .
\end{eqnarray}

In the semiclassical approximation, when loops of $\phi$ field
can be neglected and the density matrix $\rho_\phi$ is diagonal
the expectation values of the $\phi$ field modes are described,
as follows from Eqs. (\ref{eiWctp}, \ref{Gampm}), by the CTP effective action:
\begin{eqnarray}
\label{GammaCTP}
\Gamma_{CTP}[\phi^+, \phi^-] \sim S_0[\phi^+] -
S_0[\phi^-] + S_{IF}[\phi^+, \phi^-].
\end{eqnarray}
Here we suppose that an evolution of the system field becomes
semiclassical (for long wavelength modes) due to interaction
with environment (see \cite{Polarski:1995jg, Lesgourgues, Kiefer:1998qe}
and \cite{Kiefer:2008ku} with references therein).
The imaginary part of the influence action drives the system to this
semiclassical behavior, in the course of fast inflationary expansion.
The time of decoherence is proportional to $a^4g^4$ \cite{Hu:1992ig}.

In the CTP formalism, the time integration in the expressions for
actions $S$ is carried out along the closed path going from the
initial time to $+\infty$ and back. Field values generally are
not considered to be same on the forward and backward parts of the
contour, which is equivalent to doubling of degrees of freedom, i.e.,
considering two fields, $\psi^+$ and $\psi^-$. In formulas below we
will use the linear transformation (Keldysh rotation) which leads
to two new fields
$\phi_c$ and $\phi_\Delta$, defined by
\begin{equation}
\phi_c = \frac{1}{2} \left( \phi^+ + \phi^- \right) , \qquad
\phi_\Delta =  \phi^+ - \phi^- .
\end{equation}


The influence action $S_{IF}$ is calculated perturbatively
(see, e.g., \cite{Gleiser:1993ea, Lombardo:1995fg, Greiner:1996dx, Calzetta:1999zr}),
we keep the terms proportional
to $g^2$ and $g^4$. The inclusion of the terms of order $g^4$
is crucial because this is the lowest order at which an
imaginary part of the action appears. Just this part determines
the stochastic forces in the equation of motion for the system
field.
The influence action $S_{IF} \left[ \phi_c , \phi_\Delta \right]$
is given by the following formulas, separately for its real
and imaginary parts:
\begin{equation}
\label{ImdeltaAg4}
{\it Im} S_{IF} = g^4 \int d^4 x \int d^4 x'
 \phi_\Delta(x) \phi_c(x)  \phi_\Delta(x') \phi_c(x')
 {\it Re} G_{++}^{\Lambda 2}(x, x') \; ,
\end{equation}
\begin{equation}
{\it Re}  G_{++}^2 (x, x')  = - \frac{1}{2}
 \left \{ \langle \chi(x)\chi(x')\rangle^2  +
   \langle \chi(x')\chi(x)\rangle^2 \right \} \; ,
\end{equation}
\begin{eqnarray}
\label{RedeltaAg2}
{\it Re} S_{IF} = g^2 \int d^4 x \phi_c(x)  \phi_\Delta(x) i
G_{++}^{\Lambda}(x, x) a^4(\eta) -
\qquad \qquad \qquad \qquad  \qquad \qquad \qquad \qquad \\ \nonumber
- \frac{g^4}{2} \int d^4 x \int d^4 x'
\phi_\Delta(x) \phi_c(x) \left[ \phi_\Delta^2(x') + 4 \phi_c^2(x')
\right] \cdot {\it Im} G_{++}^{\Lambda 2}(x, x') \cdot \theta(\eta - \eta') \; ,
\end{eqnarray}
\begin{eqnarray}
{\it Im} G_{++}^2 (x, x')=
- \left( \theta(\eta - \eta') - \theta(\eta' - \eta) \right)
\frac{1}{2i}
\left \{ \langle \chi(x)\chi(x')\rangle^2  -  \langle \chi(x')\chi(x)\rangle^2 \right \} .
\end{eqnarray}
In these relations, the function $G_{++}(x, x')$ is the real
time propagator \cite{Chou:1984es, Landsman:1986uw} of the
$\chi$-particles on the contour,
\begin{eqnarray}
G_{++}(x, x') = i \langle T \chi(x) \chi(x') \rangle \;.
\end{eqnarray}
The upper index $\Lambda$ in the propagator symbols in Eqs. (\ref{ImdeltaAg4},
\ref{RedeltaAg2}) signifies a necessity of the cut-off in
integration over inner momenta.

A real part of a square of the propagator needed for a calculation
of the imaginary part of $S_{IF}$, can be obtained if mode functions
of the $\chi$-field, $\chi_{\vec q}(\eta, \vec x)$, are known:
\begin{eqnarray}
\label{ReGppLam2}
{\it Re} G_{++}^{\Lambda 2}(\eta, \eta', \vec k) = - (2\pi)^3
\int\limits_{q > \Lambda} d^3 q \int\limits_{q' > \Lambda} d^3 q'
\delta(\vec q + \vec q' - \vec k) \cdot
{\it Re} \left \{ \chi_q(\eta) \chi_q^*(\eta') \chi_{q'}(\eta) \chi_{q'}^*(\eta') \right \}
\; ,
\end{eqnarray}
\begin{eqnarray}
\chi_{\vec q} (\eta, \vec x) = \chi_q(\eta) e^{i \vec q \vec x} \; .
\end{eqnarray}
For the environment $\chi$-field we assume the Bunch-Davies vacuum, i.e.,
\begin{eqnarray}
\label{BDvac}
\chi_{q} (\eta) =  \frac{1}{(2\pi)^{3/2}}
\frac{e^{-i k \eta}}{a(\eta) \sqrt{2q}}
\left( 1 - \frac{i}{q \eta} \right) \; .
\end{eqnarray}
The integration in (\ref{ReGppLam2}) with using of (\ref{BDvac}) reduces
to a calculations of integrals of the form \cite{Lombardo:2005iz}:
\begin{eqnarray}
\int\limits_{q > \Lambda} d^3 q \int \limits_{q' > \Lambda}d^3 q'
\delta(\vec q + \vec q' - 2 \vec k_0) \cdot
\frac{\cos[(q+q')(\eta-\eta')]}{q^n q^m} =
\qquad\qquad\qquad\qquad\qquad\qquad\qquad\qquad\qquad\qquad\qquad\\ \nonumber
\frac{\pi}{k_0^{m+n-3}} \left [
\int \limits_{\frac{\Lambda}{k_0}+2}^{\infty} \frac{du}{u^{n-1}}
\int \limits_{u-2}^{u+2} \frac{dz}{z^{m-1}}  \cos(k_0 (\eta - \eta' ) (u+z))
+ \int\limits_{\frac{\Lambda}{k_0}}^{\frac{\Lambda}{k_0}+2} \frac{du}{u^{n-1}}
\int\limits_{\frac{\Lambda}{k_0}}^{u+2} \frac{dz}{z^{m-1}} \cos (k_0 (\eta - \eta' ) (u+z))
\right ] \; .
\end{eqnarray}

The semiclassical equation of motion for the system field is
obtained by extremizing the CTP effective action $\Gamma_{CTP}$,
\begin{eqnarray}
\label{leftdotAeff}
\left. \frac{\delta \Gamma_{CTP} }{\delta\phi_\Delta}
 \right| _ {\phi_\Delta=0} = 0 \;.
\end{eqnarray}
This is the average equation of motion, it has to be interpreted
as an average over random (stochastic) forces.  Such an
interpretation has been suggested many years ago, in studies
of quantum Brownian motion \cite{Caldeira:1982iu} by the Feynman and
Vernon method. To take these stochastic forces
into account one must keep in game the imaginary part
of the effective action (Eq. (\ref{ImdeltaAg4})) which is,
in lowest order, quadratic in $\phi_\Delta$ and, by this
reason, does not contribute to the average equation.
Standard trick for
this aim is an use of the Hubbard-Stratonovich transformation \cite{HS},
which introduces an auxiliary random field $\xi(x)$. Namely, the imaginary
part of the effective action is rewritten in the form
\begin{eqnarray}
e^{- {\it Im} S_{IF}} = \int {\cal D}\xi P[\xi]
e^{-i \int d^4 x \phi_c(x) \phi_\Delta(x) \xi(x) } \equiv
\langle e^{-i \int d^4 x \phi_c(x) \phi_\Delta(x) \xi(x) }\rangle_\xi \;.
\end{eqnarray}
Here, $P[\xi]$ is a normalized probability distribution on a
space of functions $\xi(x)$,
\begin{eqnarray}
P[\xi] = N e^{- \frac{1}{2} \int d^4 x \int d^4 x' \xi(x) \nu^{-1}(x, x') \xi(x') } \;.
\end{eqnarray}
The kernel $\nu^{-1}(x, x')$ is defined by the relation
\begin{eqnarray}
\label{nuxxp}
\nu(x, x') = g^4 a^4(\eta) a^4(\eta') {\it Re} G_{++}^{\Lambda 2}(x,x') =
\langle \xi(x) \xi(x') \rangle _\xi \;.
\end{eqnarray}
After this transformation one obtains the stochastic effective
action
\begin{eqnarray}
\Gamma_{CTP} \left[ \phi_c , \phi_\Delta \right] =
 {\it Re} \Gamma_{CTP} \left[ \phi_c , \phi_\Delta \right]  -
 \int d^4 x \phi_c(x) \phi_\Delta(x) \xi(x) \;.
\end{eqnarray}
Statistical averages are defined as functional integrals
over the $\xi(x)$-field, i.e., the averages over all realizations
of $\xi(x)$,
\begin{eqnarray}
\langle \left( ... \right) \rangle_\xi =
\int {\cal D}[\xi] {\cal P}(\xi) \left( ... \right) \; .
\end{eqnarray}
It follows from Eqs. (\ref{ImdeltaAg4}, \ref{nuxxp}) that
a correlation of the random forces is determined by an
imaginary part of the influence action.

The functional variation of the stochastic effective
action leads to the stochastic Langevin equation for the
system (inflaton) field \cite{Lee:2011fj}. Decomposing
$\phi$ on the mean field and the classical perturbation
\begin{eqnarray}
\phi(\eta, \vec x) = \phi_0(\eta) + \delta\phi(\eta, \vec x) \;,
\end{eqnarray}
one obtains, finally, the linearized Langevin equation for
the inflaton perturbation
\begin{eqnarray}
\label{deltaphipp}
\delta\phi''(\vec k, \eta) + 2 a H \delta\phi'(\vec k, \eta)
+ k^2 \delta\phi(\vec k, \eta) + a^2 m_\phi^2 \delta\phi(\vec k, \eta) =
\frac{g^2}{a^2} \phi_0 \xi \;,
\end{eqnarray}
\begin{eqnarray}
m_\phi^2 = \frac{d^2 V}{d\phi_0^2} + g^2 \langle \chi^2 \rangle \; .
\end{eqnarray}
Here, $\langle \chi^2 \rangle$-factor arises from the relation
\begin{eqnarray}
G_{++}^{\Lambda}(x, x') = i \langle T \chi^+(x) \chi^+(x') \rangle
\xrightarrow{x \to x'} i \langle \chi^2 \rangle \; .
\end{eqnarray}
The equation (\ref{deltaphipp}) contains, in its right-hand side,
the term proportional to $\xi(x)$ (this term is absent in the average
Eq. (\ref{leftdotAeff})) which is the fluctuation induced by
the (colored) stochastic noise.

Note that deriving the stochastic Langevin equation
(\ref{deltaphipp}), we neglected the dissipative term,
proportional to $g^4 {\it Im} G_{++}^{\Lambda 2}$, hoping that
it does not lead to a large error due to a small value of $g^2$.

\section{The noise-driven power spectrum of inflaton fluctuations}

The power spectrum of the quantum field fluctuations $\delta\phi$
is the function $P_\phi(k, \eta)$, which is given by the relation
\begin{eqnarray}
\langle \delta\phi(\vec x, \eta) \delta\phi(\vec x + \vec r, \eta) \rangle =
\int \frac{d^3 k}{(2\pi)^3} P_\phi(k, \eta) e^{-i \vec k \vec r} \;,
\end{eqnarray}
\begin{eqnarray}
P_\phi(k, \eta) \equiv 2 \pi^2 k^{-3} \Delta_\phi^2(k, \eta) \; .
\end{eqnarray}
The particular solution of Eq. (\ref{deltaphipp}) in a case when
$m_\phi \approx 0$ is given by the formula (see, e.g.,
\cite{Lombardo:2005iz})
\begin{eqnarray}
\label{deltaphip}
\delta\phi^{(p)}(k, \eta) = - \int\limits_{\eta_i}^{\eta} d\eta' g(k, \eta, \eta')
\xi(k, \eta') \phi_0(\eta') \; ,
\end{eqnarray}
\begin{eqnarray}
g(k, \eta, \eta') = \frac{1}{a(\eta)a(\eta')} \left[
\frac{\sin k(\eta-\eta')}{k}\cdot \left( 1 + \frac{1}{k^2\eta\eta'} \right) -
\frac{\cos k(\eta-\eta')}{k^2\eta\eta'}\left(\eta - \eta'\right)
\right] \; .
\end{eqnarray}
Using Eq. (\ref{deltaphip}), the spectrum quantity $\Delta_\phi^2(k, \eta)$
is obtained from the integral
\begin{eqnarray}
\label{frac2pi2k3Delta}
\frac{2\pi^2}{k^3} \Delta_\phi^2(k, \eta) \delta(\vec k - \vec k') =
\int\limits_{\eta_i}^{\eta} d\eta' \int\limits_{\eta_i}^{\eta} d\eta''
\phi_0(\eta') \phi_0(\eta'') \langle \xi(k, \eta') \xi(k', \eta'') \rangle_\xi
g(k, \eta, \eta') g(k', \eta, \eta'') \; .
\end{eqnarray}
The $\xi$-correlator in r.h.s. of Eq. (\ref{frac2pi2k3Delta}) is
expressed through the Fourier transform of the
$\xi$-correlator in $(\vec x, \eta)$-space (Eq. (\ref{nuxxp})),
\begin{eqnarray}
\langle \xi(k, \eta') \xi(k', \eta'') \rangle_\xi =
(2\pi)^3 \delta(\vec k - \vec k') g^4 a^4(\eta') a^4(\eta'')
{\it Re} G_{++}^{\Lambda 2}(\eta', \eta'', k) \;.
\end{eqnarray}
With an use of this equation, the noise-driven power spectrum can be expressed as:
\begin{eqnarray}
\label{Deltadeltaphi}
\Delta_{\phi}^2(k) = - \frac{g^4 k^3}{\pi^2} \int\limits_{\eta_i}^{\eta} d\eta'
\int\limits_{\eta_i}^{\eta} d\eta'' a^4(\eta') a^4(\eta'')
\phi_0(\eta') \phi_0(\eta'')
g(k, \eta, \eta') g(k, \eta, \eta'') {\it Re} G_{++}^{\Lambda 2}(\eta', \eta'', \vec k)\;.
\end{eqnarray}
The spectrum in Eq. (\ref{Deltadeltaphi}) is, in general, not
scale-invariant due to a finite duration of the inflation stage
and, also, due to existence of the (infrared) cut-off $\Lambda$.
Really, one can rewrite Eq. (\ref{Deltadeltaphi}) in a form
\cite{Lombardo:2005iz, Calzetta:1996sv}:
\begin{eqnarray}
\label{DeltadeltaphiOTHER}
\Delta_{\phi}^2(k) = - \frac{g^4}{\pi^2}
\int\limits_{k \eta_i}^{k \eta} \frac{d z'}{(z')^4}
\int\limits_{k \eta_i}^{k \eta} \frac{d z''}{(z'')^4}
\phi_0(\eta') \phi_0(\eta'')
f(k\eta, z') f(k\eta, z'') F(z', z'', \frac{\Lambda}{k})  \; ,\\
f(k\eta, z') \equiv k^3 H^{-2} g(k, \eta, \eta') =
 (z' k\eta + 1) \sin(k\eta - z') - (k\eta - z')\cos(k\eta - z') \; , \\
F(z', z'', \frac{\Lambda}{k}) \equiv k^3 H^{-4}
{\it Re} G_{++}^{\Lambda 2}(\eta', \eta'', \vec k)\;. \qquad \qquad \qquad \qquad
\end{eqnarray}

Note, once more, that this, relatively simple, spectrum formula
is obtained in a massless limit: it is assumed that both $m_\phi$
and $m_\chi$ are close to zero ($m_{\phi, \chi}^2 \ll H^2 $).
In this case, the corresponding mode functions are given by formula
(\ref{BDvac}). It was shown in \cite{Lee:2011fj} that the
approximation of massless fields is justified if $H\Delta t \cong 1$,
where $\Delta t$ is a time scale on which the particle production
happens, $\Delta t \sim (g \dot \phi)^{-1/2}$ \cite{Kofman:1997yn,
Kofman:2004yc, Green:2009ds}. The number density of the
$\chi$-particles produced is estimated as $m_\chi \langle \chi^2 \rangle$,
and, for a massless field, $\langle \chi^2 \rangle \cong H^3t/(4\pi^2)$
\cite{BDetal}. The effective mass of $\chi$ is equal to $g^2\phi_0^2$,
as follows from the coupling term in Lagrangian (\ref{calLders}),
\begin{eqnarray}
\label{mchi2}
m_\chi^2 = g^2 \phi_0^2 \approx g^2 \dot\phi^2 (t_0-t)^2.
\end{eqnarray}
Here we assume, for simplicity, that $\dot \phi$ is approximately
constant during slow-roll period of inflation.
Near the moment $t=t_0$ ($t_0$ is the time when the inflaton field
reaches the trapping point) the $\chi$-particles
are almost massless and particle production process is effective.
It follows from Eq. (\ref{mchi2}) that, if $H\Delta t \cong 1$,
$m_\chi^2 \ll H^2$ inside the time interval $\Delta t$, and
the number density of the $\chi$-particles is
$n_\chi \sim m_\chi \langle \chi^2 \rangle \sim H^3 \sim (\Delta t)^{-3}$.
The effective mass of $\phi$ is about $g^2 \langle \chi^2 \rangle$
which is also rather small due to a smallness of the coupling
constant.

It follows from these arguments that for a calculation of noise-driven power
spectrum using Eq. (\ref{Deltadeltaphi}) we must limit ourselves
(for each trapping point) by the integration over those
$\eta$'s which correspond to cosmic times $t$ close to $t_0$.
Following \cite{Lee:2011fj}, we approximate $\phi_0(\eta)$
by the relation
\begin{eqnarray}
\phi_0(\eta) = \frac{v}{H} \ln \frac{\eta}{\eta_0} \; ,
\end{eqnarray}
where $\eta_0$ is the conformal time corresponding to $t_0$, i.e.,
$\eta_0 = -1/a(t_0)H$, $v$ is the slow-roll velocity of $\phi_0$,
$v = |\dot\phi|$.
Using this approximation, integration
region in (\ref{Deltadeltaphi}) for each trapping point reduces
to the interval
\begin{eqnarray}
\eta_0 e^{H\Delta t / 2} \; < \;  \eta', \eta'' \; < \; \eta_0 e^{-H\Delta t / 2} \; .
\end{eqnarray}

\begin{figure}[!t]
\center %
\includegraphics[width=7.8 cm, trim = 0 0 0 0 ]{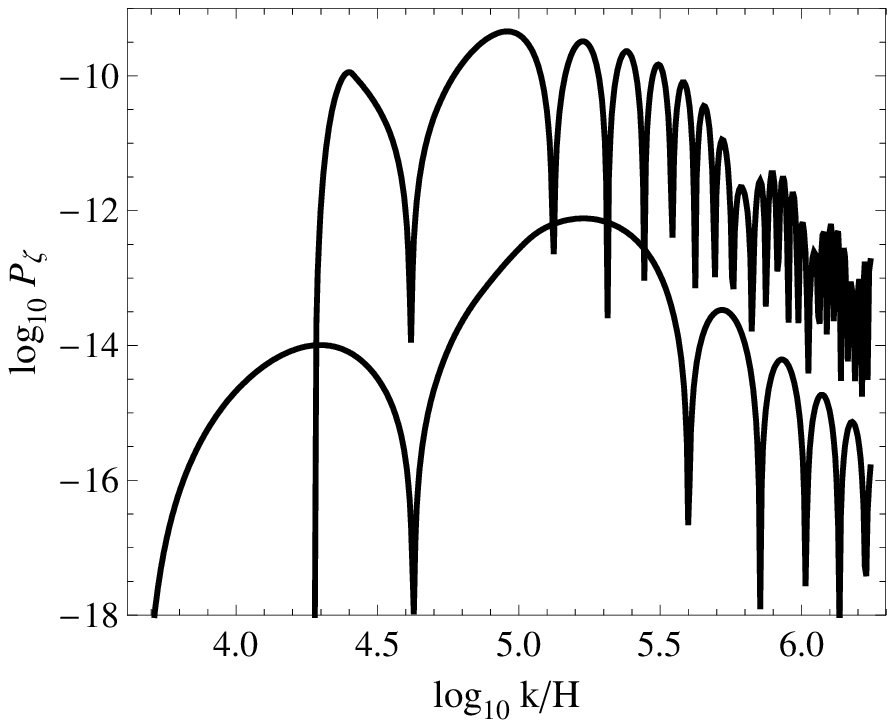}
$\quad$
\includegraphics[width=7.8 cm, trim = 0 0 0 0 ]{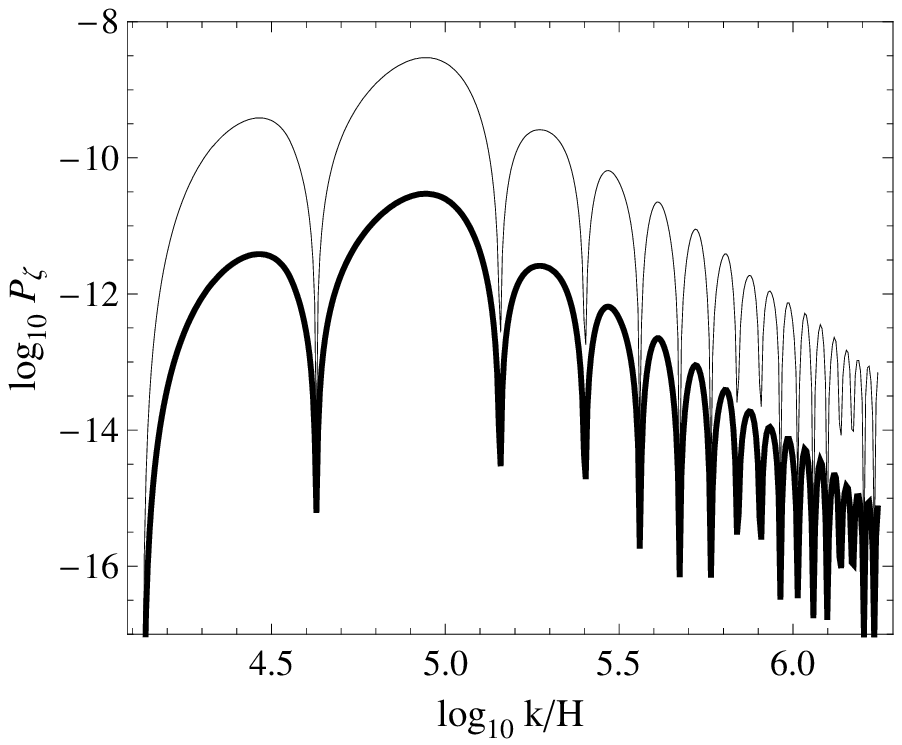}
\caption{ \label{fig-1}
Curvature perturbation power spectra generated by a single trap at $N=10$
($N$ is the e-fold number counting from the beginning of inflation).
{\bf Left panel:} $g^2=10^{-9}$ ($H\Delta t = 0.3$, upper curve);
$g^2=10^{-5}$ ($H\Delta t = 3$, lower curve).
For both curves, $\Lambda = H$.
{\bf Right panel:} $g^2=10^{-7}$ ($H\Delta t = 1$). Upper curve: $\Lambda = 0.1H$.
Lower curve: $\Lambda = H$.
}
\end{figure}

\begin{figure}
\center %
\includegraphics[width=7.8 cm, trim = 0 0 0 0 ]{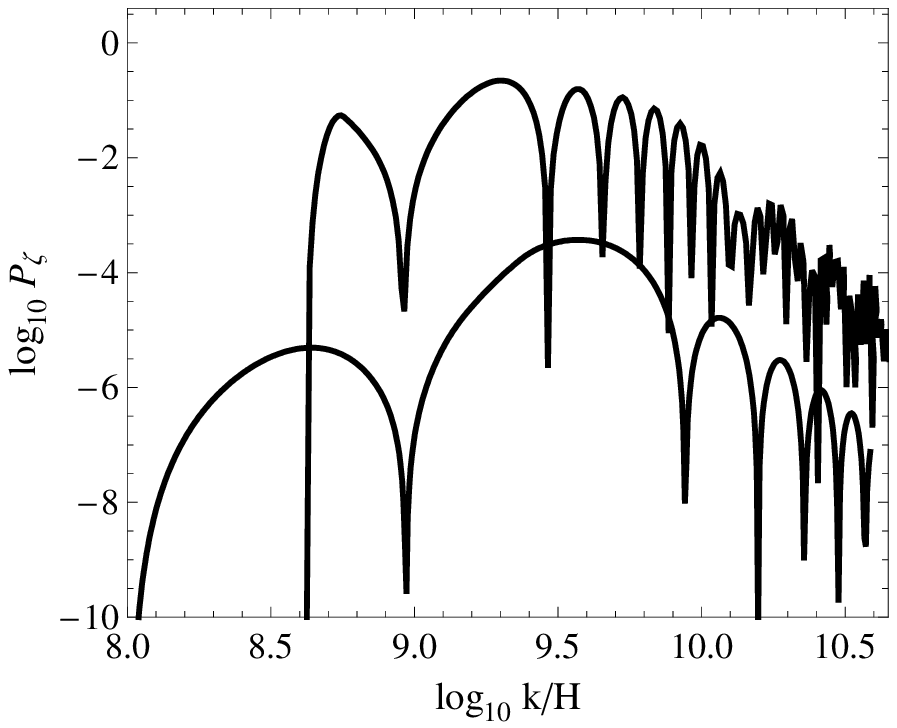}
$\quad$
\includegraphics[width=7.8 cm, trim = 0 0 0 0 ]{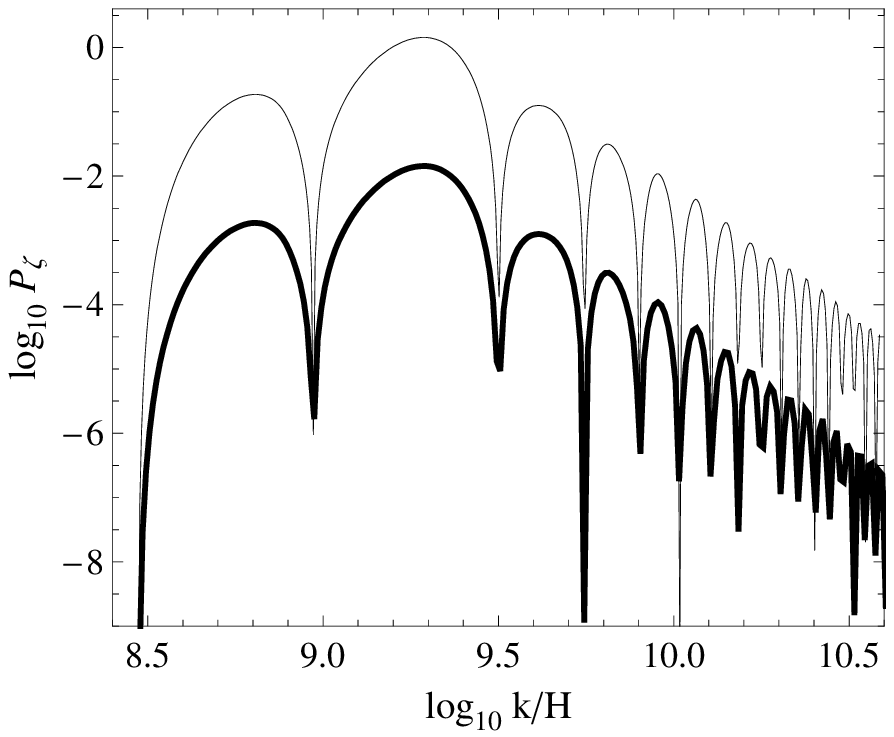}
\caption{ \label{fig-2}
Curvature perturbation power spectra generated by a single trap at $N=20$.
{\bf Left panel:} $g^2=10^{-9}$ ($H\Delta t = 0.3$, upper curve);
$g^2=10^{-5}$ ($H\Delta t = 3$, lower curve).
For both curves, $\Lambda = H$.
{\bf Right panel:} $g^2=10^{-7}$ ($H\Delta t = 1$). Upper curve: $\Lambda = 0.1H$.
Lower curve: $\Lambda = H$.
}
\end{figure}

\section{Results and discussions}

Calculating the power spectrum we are interested, mostly, in the
region of rather large comoving wave numbers, $k \gg H$.
It appears that the spectrum expression is rather
sensitive to a value of $\Lambda$, if $k \gg \Lambda$.
The cut-off value enters the spectrum expression through the
factor ${\it Re} G_{++}^{\Lambda 2}$, in which the leading term
at large ratio $k / \Lambda$ is proportional
to $(k / \Lambda)^2$ \cite{Lombardo:2005iz}:
\begin{eqnarray}
{\it Re} G_{++}^{\Lambda 2}(\eta', \eta'', \vec k) \sim \frac{H^4}{k^3}
\left( \frac{k}{2 \Lambda} \right)^2
\cos \left[ \frac{2\Lambda}{k} (k\eta' - k\eta'') \right] \;.
\end{eqnarray}

As is discussed in the Introduction (see also \cite{Vilenkin:1983xp}), infrared
cut-off is necessary to avoid an infrared singularity
in the free propagator associated with a minimally coupled massless
scalar field in de Sitter geometry. 
We assume, following \cite{Ford:1977in}, that the reasonable cut-off
value, $\Lambda$, is close to $H$.

The results of the power spectrum calculation (for one trap and
different values of $g^2$ and $\Lambda$) are shown
in Figs. \ref{fig-1} and \ref{fig-2}.
Note that the peak value of the power spectrum shifts with a change of the trap position,
$\eta_0$, in such a way that $k_{peak}\eta_0 \sim 1$. This means that the power spectrum
is formed in the near-horizon region, where the inflaton field can be considered as
classical, and, therefore, our semiclassical approach is justified.

\begin{figure}
\center %
\includegraphics[width=7.8 cm, trim = 0 0 0 0 ]{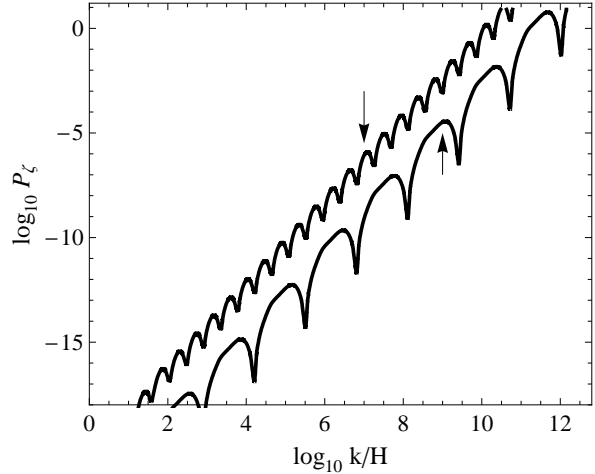}
\caption{ \label{fig-3}
Curvature perturbation power spectra generated by a series of traps.
Upper curve: $g^2=10^{-7}$ ($H\Delta t = 1$), $\Delta N = 1$.
Lower curve: $g^2=10^{-9}$ ($H\Delta t = 3$), $\Delta N = 3$.
For both cases, $\Lambda = H$.
Arrows show, for each curve, the position of $k$ at which
$g^2(k/\Lambda) = 1$.
}
\end{figure}

\begin{figure}
\center %
\includegraphics[width=7.8 cm, trim = 0 0 0 0 ]{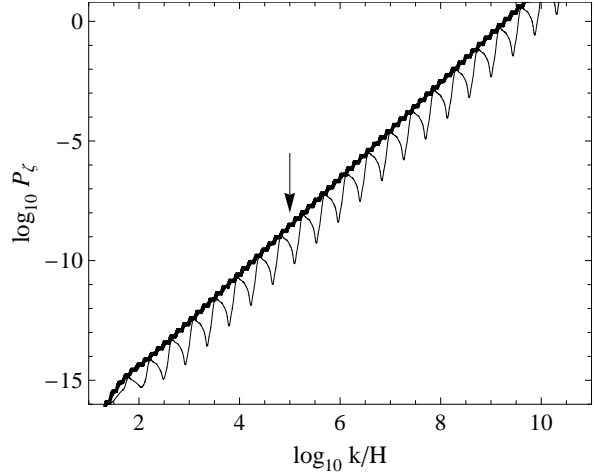}
\caption{ \label{fig-4}
Curvature perturbation power spectra generated by a series of traps,
for $g^2=10^{-5}$ ($H\Delta t = 0.3$), $\Lambda = H$.
Thin curve is for $\Delta N = 1$, thick one is for $\Delta N = 0.3$.
Arrows show the position of $k$ at which $g^2(k/\Lambda) = 1$.
}
\end{figure}

Accumulative power spectrum for a series of equally-spaced traps (with interval $\Delta N$,
in e-foldings, between them) are shown in Figs. \ref{fig-3} and \ref{fig-4}.
It is seen that the accumulative power spectrum is blue. As seen
from Fig. \ref{fig-4}, it is more or less wiggly, depending on
the relation between $\Delta N$ and $H\Delta t$ (we always
assume $\Delta N \ge H\Delta t $).

The resulting power spectrum obtained from Figs. \ref{fig-3} and \ref{fig-4}
(approximating the curves by its envelope) can be described by the following formula:
\begin{eqnarray}
\label{calPalphag2k2}
{\cal P}_\zeta(k) = \left( \frac{H}{v} \right)^2 \Delta_\phi^2
 \approx \alpha \; g^2 \left( \frac{k}{H} \right)^2 \;, \quad
\alpha \approx 5\times 10^{-14}.
\end{eqnarray}
This form of the $k$-dependence of ${\cal P}_\zeta(k)$ is similar
with analogous dependence predicted in \cite{Lee:2011fj}
(see \cite{Lin:2012gs} where the numerical results of \cite{Lee:2011fj}
had been parameterized).

Now it is convenient to rescale the scale factor and, correspondingly,
the comoving wave number setting $a=1$ at present time (rather than
$a=1$ at the initial time of the inflation era, as was set before).
After this rescaling the Eq. (\ref{calPalphag2k2}) is rewritten as
\begin{eqnarray}
\label{calPalphag2k2scaled}
{\cal P}_\zeta(k) \approx \alpha \; g^2 \left( \frac{k}{H a_{start}} \right)^2 \;, \quad
\alpha \approx 5\times 10^{-14},
\end{eqnarray}
where $a_{start}$ is a scale factor at the beginning of inflation. It is related
to a scale factor at the end of inflation, $a_{end}$, by the formula
\begin{eqnarray}
\frac{a_{end}}{a_{start}} = e^{N_{inf}},
\end{eqnarray}
where $N_{inf}$ is the total number of e-folds during inflation.
The value of $a_{end}$ can be easily estimated by the approximate
equation,
\begin{eqnarray}
a_{end} = a_{eq} \frac{T_{eq}} {\sqrt{M_{Pl} H} } \;,
\end{eqnarray}
where $a_{eq}$ and $T_{eq}$ are the scale factor and temperature
of the Universe at the time of the matter-radiation equality
($T_{eq} \sim 3  {\rm eV}$). As a result, the spectrum amplitude
contains the factor $\exp(2 N_{infl})/H$. Observations at
cosmologically large scales ($k \lesssim 10 \; {\rm Mpc}^{-1}$)
show (see, e.g., \cite{Ade:2015xua}) that the amplitude of the primary power
spectrum does not exceed $\sim 10^{-9}$. To match this condition,
the duration of inflation in our model must be not too long
($N_{inf} \lesssim 70 \div 75$) and the energy scale of inflation
must be rather high ($H \gtrsim 10^8 \div 10^{10} \; {\rm GeV}$).

It is important to check when (at which values of $k$ and/or PBH mass)
the resulting power spectrum (\ref{calPalphag2k2scaled})
reaches values that are prohibited by the PBH overproduction. In this work we will
assume that, roughly, PBHs are over-produced when, for some value of $k$,
${\cal P}_\zeta(k) > {\cal P}_\zeta^{PBH}$ where ${\cal P}_\zeta^{PBH} \sim 10^{-2}$
is the PBH production threshold (see, e.g., \cite{Carr:2009jm, Bugaev:2010bb}).
Then, having the relation between $k$ and horizon mass \cite{Bugaev:2010bb},
\begin{eqnarray}
k \approx 2\times 10^{23} (M_h[{\rm g}])^{-1/2} {\rm Mpc}^{-1} \;,
\end{eqnarray}
and, assuming that, approximately, the PBH mass is of order
of horizon mass, $M_{BH} \approx M_h(k)$, we obtain, from Eq.
(\ref{calPalphag2k2scaled}), the corresponding border value of $M_{BH}$,
\begin{eqnarray}
M_{BH}^{(b)} \approx \frac{3\times 10^{14} {\rm GeV}}{H} \cdot e^{2N_{inf}}
 \cdot \frac{\alpha g^2}{{\cal P}_\zeta^{PBH}} \; {\rm g} \;.
\end{eqnarray}

Clearly, the production of PBHs with masses $M_{BH} < M_{BH}^{(b)}$ is prohibited by the
present constraints \cite{Carr:2009jm, Bugaev:2008gw, Josan:2009qn, Bugaev:2010bb}
if the power spectrum grows with $k$ as strongly as
Eq. (\ref{calPalphag2k2scaled}) predicts.

\begin{figure}
\center %
\includegraphics[width=7.8 cm, trim = 0 0 0 0 ]{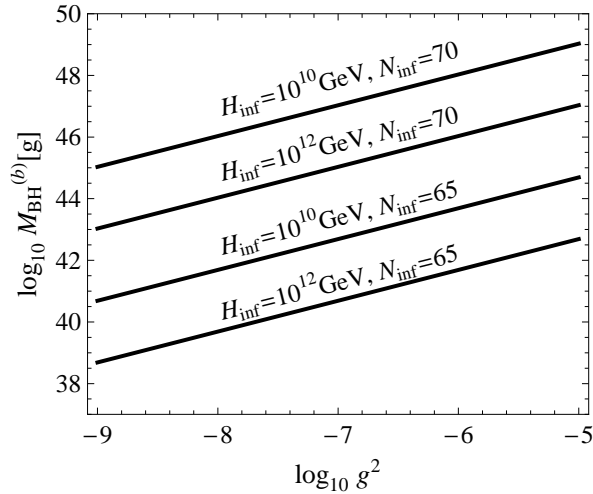}
\caption{ \label{fig-5}
The values of border masses of PBHs for different values
of model parameters.
}
\end{figure}

The resulting dependence of $M_{BH}^{(b)}$ on $g^2$, $N_{inf}$ and $H$ is shown
in Fig. \ref{fig-5}. It is seen that the inadmissibly large
PBH overproduction in the inflation model considered in the present paper is
predicted for all reasonable values of parameters $N_{inf}$,
$H$ and for some interval of values of the coupling constant $g^2$.
This is the main conclusion of the paper.

The model of inflation with trapping points (in a particular case
of the weak  coupling constant $g^2$) can survive
if, by some reason, the growth of the power spectrum with increase of $k$ becomes gradually
more slow and, finally, stops at some value ${\cal P}_{\zeta, max} < 10^{-2}$.
One can imagine two reasons, at least, for such a behavior:
{\it i)} an accounting of the dissipation term
in the influence action $S_{IF}$ will, in general, lead to a damping of the power of
high-$k$ modes \cite{Lee:2011fj} and {\it ii)} an accounting
of higher order terms in the perturbative
expansion (these terms correspond to diagrams with two or more $\chi$-loops) also can
change a form of the spectrum. Our calculations based on the perturbation theory are
reliable if $g^2 (k/\Lambda) \lesssim 1$ (the corresponding value of $k$ is denoted by
arrow in Figs. \ref{fig-3} and \ref{fig-4}). So, parts of spectrum curves in these figures
which are to the right of the arrows have to be considered as extrapolations (hoping
that the accounting of higher order terms in $g^2$ doesn't slow down the growth of
the spectrum).

Note, in the end, that our numerical results for the power spectrum depend rather strongly
on a choice of the infra-red cut-off value $\Lambda$
(see Figs. \ref{fig-1} and \ref{fig-2}).
Another thing which is worth mentioning: power spectra from individual trapping points
are overlapped weakly so the envelope curve of the accumulative spectrum almost does not
depend on the spacing (see Figs. \ref{fig-3} and \ref{fig-4}).

\section*{Acknowledgments}
The work of P.K. was partially supported by the grant of President of RF number SS-3110.2014.2.


\begin{thebibliography}{00}  

\bibitem{Kofman:1986wm}
  L.~A.~Kofman and A.~D.~Linde,
  Nucl.\ Phys.\ B {\bf 282}, 555 (1987).

\bibitem{Chung:1999ve}
  D.~J.~H.~Chung, E.~W.~Kolb, A.~Riotto and I.~I.~Tkachev,
  Phys.\ Rev.\ D {\bf 62}, 043508 (2000)
  [hep-ph/9910437].

\bibitem{Kofman:2004yc}
  L.~Kofman, A.~D.~Linde, X.~Liu, A.~Maloney, L.~McAllister and E.~Silverstein,
  JHEP {\bf 0405}, 030 (2004)
  [hep-th/0403001].

\bibitem{Romano:2008rr}
  A.~E.~Romano and M.~Sasaki,
  Phys.\ Rev.\ D {\bf 78}, 103522 (2008)
  [arXiv:0809.5142 [gr-qc]].

\bibitem{Green:2009ds}
  D.~Green, B.~Horn, L.~Senatore and E.~Silverstein,
  Phys.\ Rev.\ D {\bf 80}, 063533 (2009)
  [arXiv:0902.1006 [hep-th]].

\bibitem{Barnaby:2009mc}
  N.~Barnaby, Z.~Huang, L.~Kofman and D.~Pogosyan,
  Phys.\ Rev.\ D {\bf 80}, 043501 (2009)
  [arXiv:0902.0615 [hep-th]].

\bibitem{Barnaby:2010ke}
  N.~Barnaby,
  Phys.\ Rev.\ D {\bf 82}, 106009 (2010)
  [arXiv:1006.4615 [astro-ph.CO]].

\bibitem{Lee:2011fj}
  W.~Lee, K.~W.~Ng, I.~C.~Wang and C.~H.~Wu,
  Phys.\ Rev.\ D {\bf 84}, 063527 (2011)
  [arXiv:1101.4493 [hep-th]].

\bibitem{Wu:2006xp}
  C.~H.~Wu, K.~W.~Ng, W.~Lee, D.~S.~Lee and Y.~Y.~Charng,
  JCAP {\bf 0702}, 006 (2007)
  [astro-ph/0604292].

\bibitem{Kofman:1994rk}
  L.~Kofman, A.~D.~Linde and A.~A.~Starobinsky,
  Phys.\ Rev.\ Lett.\  {\bf 73}, 3195 (1994)
  [hep-th/9405187].

\bibitem{Kofman:1997yn}
  L.~Kofman, A.~D.~Linde and A.~A.~Starobinsky,
  Phys.\ Rev.\ D {\bf 56}, 3258 (1997)
  [hep-ph/9704452].

\bibitem{Seery:2008qj}
  D.~Seery, K.~A.~Malik and D.~H.~Lyth,
  JCAP {\bf 0803}, 014 (2008)
  [arXiv:0802.0588 [astro-ph]].

\bibitem{Liu:2014ifa}
  G.~C.~Liu, K.~W.~Ng and I.~C.~Wang,
  Phys.\ Rev.\ D {\bf 90}, no. 10, 103531 (2014)
  [arXiv:1409.3661 [hep-ph]].

\bibitem{Morikawa:1989xz}
  M.~Morikawa,
  Phys.\ Rev.\ D {\bf 42}, 1027 (1990).

\bibitem{HZ}
  B. L.~Hu and Y.~Zhang, Coarse-Graining, Scaling, and Inflation (Univ.
  Maryland Preprint 90-186);
  B.~L.~Hu, in Relativity and Gravitation: Classical and Quantum,
  Proc. SILARG VII, Cocoyoc, Mexico 1990. eds. J.~C.~D'Olivo et al
  (World Scientific, Singapore 1991).

\bibitem{Matarrese:2003ye}
  S.~Matarrese, M.~A.~Musso and A.~Riotto,
  JCAP {\bf 0405}, 008 (2004)
  [hep-th/0311059].

\bibitem{AAS}
  A.~A.~Starobinsky, in {\it Fundamental Interactions}, edited by V.~N.~Ponomarev (MGPI Press, Moscow, 1984), p. 54;
  A.~A.~Starobinsky, in {\it Lecture Notes in Physics Vol. 246}, edited by H.~J.~de Vega and N.~Sanchez (Springer, New York, 1986), p. 107.

\bibitem{SK}
   J.~Schwinger, Jour. Math. Phys. (N.Y.) {\bf 2}, 407 (1961);
   P~ M.~Bakshi and K.~T.~Mahanthappa, Jour. Math. Phys. (N.Y.) {\bf 4}, 1 (1963); {\bf 4}, 12 (1963);
   L. V. Keldysh, Zh. Eksp. Teor. Fiz. {\bf 47}, 1515 (1964).

\bibitem{CH}
  E.~Calzetta and B.~L.~Hu,
  Phys.\ Rev.\ D {\bf 35}, 495 (1987);
  Phys.\ Rev.\ D {\bf 37}, 2878 (1988);
  Phys.\ Rev.\ D {\bf 40}, 656 (1989).

\bibitem{Jordan:1986ug}
  R.~D.~Jordan,
  Phys.\ Rev.\ D {\bf 33}, 444 (1986).

\bibitem{Calzetta:1993qe}
  E.~Calzetta and B.~L.~Hu,
  Phys.\ Rev.\ D {\bf 49}, 6636 (1994)
  [gr-qc/9312036].

\bibitem{Hu:1994iw}
  B.~L.~Hu,
  In Banff 1993, Proceedings, Thermal field theories and their applications 309-346.
  [gr-qc/9403061].

\bibitem{FV}
  R.~Feynman and F.~Vernon, Ann. Phys. (NY) {\bf 24}, 118 (1963);
  R.~Feynman and A.~Hibbs, Quantum Mechanics and Path Integrals
  (McGraw - Hill, New York, 1965);
  H.~Kleinert. Path Integrals in Quantum Mechanics, Statistics, and Polymer
  Physics (World Scientific, Singapore, 1990).

\bibitem{Caldeira:1982iu}
  A.~O.~Caldeira and A.~J.~Leggett,
  Physica {\bf 121A}, 587 (1983); Ann.~Phys. {\bf 149}, 374 (1983).

\bibitem{Grabert:1988yt}
  H.~Grabert, P.~Schramm and G.~L.~Ingold,
  Phys.\ Rept.\  {\bf 168}, 115 (1988).
  doi:10.1016/0370-1573(88)90023-3

\bibitem{Hu:1992ig}
  B.~L.~Hu, J.~P.~Paz and Y.~Zhang,
  ``Quantum origin of noise and fluctuations in cosmology'',
  in Chateau du Pont d'Oye 1992, Proceedings, The origin of structure
  in the universe, p. 227-251
  [gr-qc/9512049].

\bibitem{Gleiser:1993ea}
  M.~Gleiser and R.~O.~Ramos,
  Phys.\ Rev.\ D {\bf 50}, 2441 (1994) [hep-ph/9311278].

\bibitem{Berera:1998gx}
  A.~Berera, M.~Gleiser and R.~O.~Ramos,
  Phys.\ Rev.\ D {\bf 58}, 123508 (1998)
  [hep-ph/9803394].

\bibitem{Yokoyama:2004pf}
  J.~Yokoyama,
  Phys.\ Rev.\ D {\bf 70}, 103511 (2004)
  [hep-ph/0406072].

\bibitem{Ramos:2013nsa}
  R.~O.~Ramos and L.~A.~da Silva,
  JCAP {\bf 1303}, 032 (2013)
  [arXiv:1302.3544 [astro-ph.CO]].

\bibitem{Boyanovsky}
  D.~Boyanovsky,
  New J.\ Phys.\  {\bf 17}, no. 6, 063017 (2015) [arXiv:1503.00156 [hep-ph]];
  Phys.\ Rev.\ D {\bf 92}, no. 2, 023527 (2015) [arXiv:1506.07395 [astro-ph.CO]];
  Phys.\ Rev.\ D {\bf 93}, 043501 (2016) [arXiv:1511.06649 [astro-ph.CO]].

\bibitem{Boyanovsky:2012}
  D.~Boyanovsky,
  Phys.\ Rev.\ D {\bf 85}, 123525 (2012) [arXiv:1203.3903 [hep-ph]];
  Phys.\ Rev.\ D {\bf 86}, 023509 (2012) [arXiv:1205.3761 [astro-ph.CO]].

\bibitem{Allen:1985ux}
  B.~Allen,
  Phys.\ Rev.\ D {\bf 32}, 3136 (1985).

\bibitem{Allen:1987tz}
  B.~Allen and A.~Folacci,
  Phys.\ Rev.\ D {\bf 35}, 3771 (1987).

\bibitem{BDetal}
  T.~S.~Bunch, P.~C.~W.~Davies, Proc. R. Soc. A360, 117 (1978);
  L.H.Ford, A.Vilenkin, Phys. Rev. {\bf D 26}, 1231 (1982);
  A.Linde, Phys. Lett. {\bf B 116}, 335 (1982);
  A.A.Starobinsky, Phys. Lett. {\bf B 117}, 175 (1982).

\bibitem{Ford:1977in}
  L.~H.~Ford and L.~Parker,
  Phys.\ Rev.\ D {\bf 16}, 245 (1977).

\bibitem{Lyth:2007jh}
  D.~H.~Lyth,
  JCAP {\bf 0712}, 016 (2007)
  [arXiv:0707.0361 [astro-ph]].

\bibitem{Bartolo:2007ti}
  N.~Bartolo, S.~Matarrese, M.~Pietroni, A.~Riotto and D.~Seery,
  JCAP {\bf 0801}, 015 (2008)
  [arXiv:0711.4263 [astro-ph]].

\bibitem{Prokopec:2007ak}
  T.~Prokopec, N.~C.~Tsamis and R.~P.~Woodard,
  Annals Phys.\  {\bf 323}, 1324 (2008) [arXiv:0707.0847 [gr-qc]].

\bibitem{Xue:2011hm}
  W.~Xue, K.~Dasgupta and R.~Brandenberger,
  Phys.\ Rev.\ D {\bf 83}, 083520 (2011)
  [arXiv:1103.0285 [hep-th]].

\bibitem{Larjo:2011uh}
  K.~Larjo and D.~A.~Lowe,
  Phys.\ Rev.\ D {\bf 85}, 043528 (2012)
  [arXiv:1112.5425 [hep-th]].

\bibitem{Polarski:1995jg}
  D.~Polarski and A.~A.~Starobinsky,
  Class.\ Quant.\ Grav.\  {\bf 13}, 377 (1996)
  [gr-qc/9504030].

\bibitem{Lesgourgues}
  J.~Lesgourgues, D.~Polarski and A.~A.~Starobinsky,
  Nucl.\ Phys.\ B {\bf 497}, 479 (1997); Class.\ Quant.\ Grav.\  {\bf 14}, 881 (1997).

\bibitem{Kiefer:1998qe}
  C.~Kiefer, D.~Polarski and A.~A.~Starobinsky,
  Int.\ J.\ Mod.\ Phys.\ D {\bf 7}, 455 (1998)
  [gr-qc/9802003].

\bibitem{Kiefer:2008ku}
  C.~Kiefer and D.~Polarski,
  Adv.\ Sci.\ Lett.\  {\bf 2}, 164 (2009)
  [arXiv:0810.0087 [astro-ph]].

\bibitem{Lombardo:1995fg}
  F.~Lombardo and F.~D.~Mazzitelli,
  Phys.\ Rev.\ D {\bf 53}, 2001 (1996)[hep-th/9508052].

\bibitem{Greiner:1996dx}
  C.~Greiner and B.~Muller,
  Phys.\ Rev.\ D {\bf 55}, 1026 (1997) [hep-th/9605048].

\bibitem{Calzetta:1999zr}
  E.~A.~Calzetta, B.~L.~Hu and F.~D.~Mazzitelli,
  Phys.\ Rept.\  {\bf 352}, 459 (2001) [hep-th/0102199].


\bibitem{Chou:1984es}
  K.~c.~Chou, Z.~b.~Su, B.~l.~Hao and L.~Yu,
  Phys.\ Rept.\  {\bf 118}, 1 (1985).

\bibitem{Landsman:1986uw}
  N.~P.~Landsman and C.~G.~van Weert,
  Phys.\ Rept.\  {\bf 145}, 141 (1987).

\bibitem{Lombardo:2005iz}
  F.~C.~Lombardo and D.~Lopez Nacir,
  Phys.\ Rev.\ D {\bf 72}, 063506 (2005) [gr-qc/0506051].

\bibitem{HS}
  R.~L.~Stratonovich, Soviet Physics Doklady {\bf 2}, 416 (1958);
  J.~Hubbard, Physical Review Letters {\bf 3}, 77 (1959).

\bibitem{Calzetta:1996sv}
  E.~A.~Calzetta and S.~Gonorazky,
  Phys.\ Rev.\ D {\bf 55}, 1812 (1997) [gr-qc/9608057].

\bibitem{Vilenkin:1983xp}
  A.~Vilenkin,
  Nucl.\ Phys.\ B {\bf 226}, 527 (1983).

\bibitem{Lin:2012gs}
  C.~M.~Lin and K.~W.~Ng,
  Phys.\ Lett.\ B {\bf 718}, 1181 (2013) [arXiv:1206.1685 [hep-ph]].

\bibitem{Ade:2015xua}
  P.~A.~R.~Ade {\it et al.} [Planck Collaboration],
  arXiv:1502.01589 [astro-ph.CO].

\bibitem{Carr:2009jm}
  B.~J.~Carr, K.~Kohri, Y.~Sendouda and J.~Yokoyama,
  Phys.\ Rev.\ D {\bf 81}, 104019 (2010)  [arXiv:0912.5297 [astro-ph.CO]].

\bibitem{Bugaev:2010bb}
  E.~Bugaev and P.~Klimai,
  Phys.\ Rev.\ D {\bf 83}, 083521 (2011) [arXiv:1012.4697 [astro-ph.CO]].

\bibitem{Bugaev:2008gw}
  E.~Bugaev and P.~Klimai,
  Phys.\ Rev.\ D {\bf 79}, 103511 (2009) [arXiv:0812.4247 [astro-ph]].

\bibitem{Josan:2009qn}
  A.~S.~Josan, A.~M.~Green and K.~A.~Malik,
  Phys.\ Rev.\ D {\bf 79}, 103520 (2009) [arXiv:0903.3184 [astro-ph.CO]].

\end{thebibliography}
\end{document}